\documentclass[12pt,onecolumn]{IEEEtran}
\IEEEoverridecommandlockouts
\usepackage{cite}
\usepackage{amsmath,amssymb,amsfonts}
\usepackage{algorithmic}
\usepackage{graphicx}
\usepackage{authblk}
\usepackage{textcomp}
\usepackage{xcolor}
\def\BibTeX{{\rm B\kern-.05em{\sc i\kern-.025em b}\kern-.08em
    T\kern-.1667em\lower.7ex\hbox{E}\kern-.125emX}}

\mathchardef\mhyphen="2D
\usepackage{url}

\newcommand*{\affaddr}[1]{#1} 
\newcommand*{\affmark}[1][*]{\textsuperscript{#1}}
\newcommand*{\email}[1]{\textit{#1}}

\begin{document}
%
\title{Cyber-Attack Consequence Prediction}

\author{%
Prerit Datta\affmark[1], Natalie Lodinger\affmark[2], Akbar Siami Namin\affmark[1], and Keith S. Jones\affmark[2]\\
\affaddr{\affmark[1]Department of Computer Science, } 
\affaddr{\affmark[2]Department of Psychological Sciences}\\
\affaddr{\affmark[1,2]Texas Tech University} \\
\email{\{prerit.datta, natalie.lodinger, akbar.namin, keith.s.jones\}@ttu.edu} 
}

\maketitle

\begin{abstract}
Cyber-physical systems posit a complex number of security challenges due to interconnection of heterogeneous devices having limited processing, communication, and power capabilities. Additionally, the conglomeration of both physical and cyber-space further makes it difficult to devise a single security plan spanning both these spaces. Cyber-security researchers are often overloaded with a variety of cyber-alerts on a daily basis many of which turn out to be false positives. In this paper, we use machine learning and natural language processing techniques to predict the consequences of cyberattacks. The idea is to enable security researchers to have tools at their disposal that makes it easier to communicate the attack consequences with various stakeholders who may have little to no cybersecurity expertise. Additionally, with the proposed approach researchers' cognitive load can be reduced by automatically predicting the consequences of attacks in case new attacks are discovered. We compare the performance through various machine learning models employing word vectors obtained using both tf-idf and Doc2Vec models. In our experiments, an accuracy of 60\% was obtained using tf-idf features and 57\% using Doc2Vec method for models based on LinearSVC model. 
\end{abstract}


\IEEEpeerreviewmaketitle

\section{Introduction}
The National Institute of Standards and Technology (NIST) defines Cyber-Physical systems (CPS) as systems ``comprising interactions between digital, analog, physical and human components'' through physics and logic for enabling smart services and improving the quality of life \cite{nist}. The concept of Cyber-physical systems can be applied to a variety of areas such as manufacturing, healthcare, agriculture, aviation, business etc. Some of the popular CPS technologies include smart grids, smart cities, Internet-of-Things (IoT) and industrial control systems. 
Due to their combination of cyber as well as the physical domains, CPS systems brings about unique challenges from both domains. In addition to physical attacks, cyberattacks remain one of the critical challenges of cyber-physical systems security. With the recent increase in cyber incidents involving CPS \cite{walker2020threats, AlMhiqani2018}, reliability and availability of CPS systems remains a top security goal \cite{serpanos2018cyber, YAACOUB2020103201}. 

Given the ever-evolving threat landscape, security researchers managing the security operation center (SOC) are often overloaded with numerous security incidents and, at the same time, trying to keep abreast with the latest threats in the wild. Situation awareness (SA) \cite{Endsley1988} is the concept of perceiving the elements in the environment, comprehending their meaning and making decisions or taking action. Cyber-situation awareness (cyber-SA) is the concept of situation awareness applied to the cyber-security domain. Cyber-SA can help security researchers to reduce their cognitive load and help them to focus on what is important for cyber threat analysis and mitigation.

Threat mitigation and analysis involves communication with different stakeholders who may not be well versed with the concepts of cybersecurity. One of the ways that cybersecurity personnel can communicate with stakeholders is through the realization of the (non)-technical consequences of attacks to end users and thus informing them about the impact of such attacks to them. 

This paper is a first step towards reducing cognitive workload of security experts and even average Internet users. The research presented in this paper investigates whether it is possible to train a model that predicts the technical and non-technical layman consequences of novel cyber attacks. The trained model should be able to digest textual descriptions of new cyber attacks through vectorization and then map them to known cyber attacks with clear consequences to end users. Machine learning-based algorithms can perform document embedding and then vectorization. More specifically, the distance of embedded vectors and a threshold value can decide whether two cyber attacks have similar semantics and thus consequences. 

We have created our own dataset of cyber attacks along with their technical and non-technical consequences. The repository consists of 93 diverse attacks and their descriptions, which are annotated with their consequences, all in textual and non-structured formats. We then apply natural language processing and machine-learning algorithms to cyber attacks descriptions to predict the consequences of the attack. 
This paper makes the following key contributions:
\begin{itemize}
    \item[--] Introduce a dataset of cyber attacks and their  (non)-technical consequences to end users. 
    \item[--] Build machine learning models to predict the (non)-technical consequences of attacks.
    \item[--] Report the performance of prediction models through experimental studies. 
\end{itemize}

This paper is organized as follows. Section \ref{sec:background} describes situation awareness and its relationship to cyber-security for CPS. Section \ref{sec:relatedwork} discusses the related work. The technical background of the employed machine learning classifiers is presented in Section \ref{sec:technical}. We describe our methodology in Section \ref{sec:methodolgy} and results in Section \ref{sec:results}. The conclusion and future work are described in Section \ref{sec:conclusion}.

\section{Situation Awareness and Cyber-Security}
\label{sec:background}
In their seminal work, Endsley \cite{Endsley1988} described the concept of situation awareness as ``the \textit{perception} of the elements in the environment within a volume of time and space, the \textit{comprehension} of their meaning and the \textit{projection} of their status in the near future''. The definition can be broken down into three components namely: perception, comprehension and projection. For Cyber-SA, \textit{perception} involves gathering  data from various threat sources, \textit{comprehending} the disparate data by deriving patterns or knowledge and then taking actions based on the implications of the decisions made (\textit{projection}). Given the voluminous amount of threat data and alerts, security practitioners often need automated tools to reduce their cognitive loads, prioritize alerts, do automated analysis, and  generate reports for decision making. Additionally, cyber-situation tools can be used by defenders and security professionals to assess decision making and decide a recourse of action \cite{Gutzwiller}.

\begin{figure}
    \centering
    \includegraphics[width=0.35\textwidth]{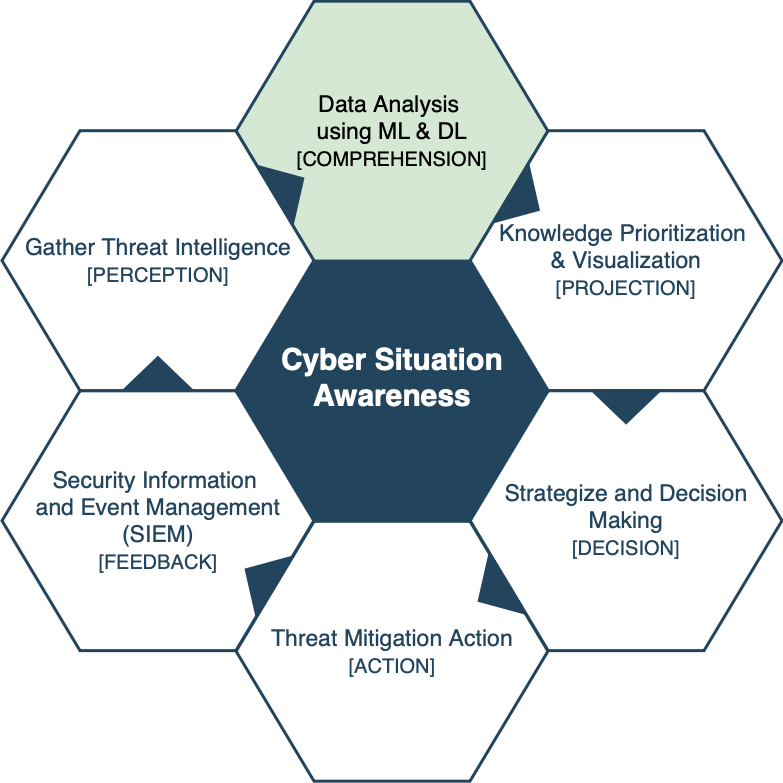}
    \caption{Cyber-Situation awareness model based on Endsley's SA model.}
    \label{fig:cyberSA}
\end{figure}

Figure \ref{fig:cyberSA} depicts the dimensions and a model for cyber-SA. The process begins by gathering the threat intelligence from various sources including the current state of sensors in the CPS system in addition to incident reports. After gathering the data, it is important that the SOC is powered with automated tools at their disposal that can use state-of-the art machine and deep learning algorithms to analyse the data. This is highlighted in green in the figure, as this is the main emphasis of the paper. After analysing the data, the SOC can prioritize the gathered knowledge and may visualize the threat reports to further strategize and make decisions after discussion with various stakeholders. After taking appropriate actions, the event logs and detailed action reports are stored in the security information and event management (SIEM) for future reference and decision-making.

\section{Related Work}
\label{sec:relatedwork}
Cyber-physical systems present different security challenges compared to traditional IT systems for a variety of reasons. The microcosm of heterogeneous devices, which have limited processing and communication abilities, makes it difficult to offer encryption during communication \cite{walker2020threats}. Additionally, establishing trust to enable communication is another important challenge for securing CPS. Attacks on CPS systems, such as manufacturing and power grids, can incur significant damages and even loss of life \cite{AlMhiqani2018}. An analysis of recent cyberattacks on CPS indicates that most of the attacks are driven to cause disruption, and are often politically motivated \cite{AlMhiqani2018, YAACOUB2020103201}.

There have been several efforts to incorporate cyber-situation awareness into CPS. Cyber-SA can be useful in simulating real cyber-incidents for decision making and training cyber defenders.  Debatty and Mees \cite{Debatty} propose a tool assessing the cyber-situation awareness for cyber-defenders in the military. The tool consists of several scenarios that can be used to train and evaluate cyber defenders by simulating cyberattack scenarios to improve skills and knowledge of the cybersecurity personnel. 

Preden \cite{Preden} proposes a middle-ware architecture to enable situation awareness in cyber-physical systems. The key idea is to have a middle-ware that proactively handles tasks like service discovery, delivery and validation of data contracts to enable communication and resource constraint satisfaction. The authors argues that this aspect of decoupling data processing from communication will allow CPS to enable Cyber-SA. 

Yang et al. \cite{Yang} propose a framework to enable Cyber-SA in a smart metering system. The authors recommend coupling of both cyber and physical attacks data and then using a knowledge model for data analysis and decision-making.  The knowledge model uses fuzzy logic to prioritize and assess attacks based on their severity, occurrence and detection using the underlying knowledge base and data from the previous step. This improves attack comprehension and helps security researchers automate the threat analysis process and make informed decisions. 

Orojloo and Azgomi \cite{Orojloo} propose a methodology to predict attackers' behavior and consequences of attacks on CPS. The authors use fuzzy logic to evaluate the possibilities of various attack scenarios with the help of an attack tree. The attacker's skills, knowledge and access are used to determine the likely attack paths with the help of expert knowledge. Attack impact is measured in terms of time-to-shutdown (TTSD) for each element in the CPS system and the attack's probability. The defenders can devise and assess the possible mitigation of various attack scenarios using this information.

Cardenas et al. \cite{Cardenas} argue that, in addition to the challenges posed by the CPS, the research community has overlooked the consequences of the attacks on the CPS infrastructure. The authors contend that while security incident reports show some of the effects of cyberattacks, the actual consequence of a successful attack is the missing link in the CPS security literature. Our work presented in this paper is an attempt to bridge this gap and open up new research directions for the cybersecurity community to focus on the consequences of cyberattacks when deciding on the security of a system.
 
\section{Technical Background}
\label{sec:technical}
\subsection{Machine Learning Models}
This section briefly discusses the technical background of the machine learning classifiers studied. 
\begin{itemize}
    \item \textbf{Linear Support Vector Classifier (SVC)}  uses a linear kernel to perform classification. LinearSVC is preferred as it allows for additional parameters for tuning performance such as penalties and loss functions. Additionally, Linear SVC allows for a one-vs-rest classification strategy to be easily applied to multi-class problems.
    
    \item \textbf{Logistic Regression} is one of the most popular models in statistics and machine learning. Inherently, logistic regression is a predictive model that identifies the relationship between features (independent variables) and the target variable (dependent variable) in terms of probability.
    
    \item \textbf{Multinomial Naive Bayes (NB)} is a classifier that is best suited for discrete features. As a result, it is well suited for text-classification problems that use tf-idf features for text representation.
    
    \item \textbf{Random Forest} classifiers use an ensemble of decision trees wherein each individual tree predicts the class output. The class having the most votes is used as the final result. These trees are collectively called a forest. The word random in the decision represent that each tree picks only a subset of available features to determine the split and thus results in low correlation between different trees in the forest.
    
    \item \textbf{Multilayer Perceptron (MLP)} is the simplest form of artificial neural network. It consists of three layers: input layer, hidden layer and the output layer. The model tries to learn patterns in the data and then after being trained on the examples, it can classify the new instances. We used the MLP classifier from the scikit-learn library with maximum iterations of 1000 and default number of hidden layers.
\end{itemize}

\subsection{tf-idf}
One of the classical NLP-based algorithms called tf-idf is used to convert text into feature vectors that can be used by machine learning algorithms. tf-idf is an acronym formed by combining two words \textit{term frequency (TF)} and \textit{inverse-document frequency (IDF)}. The \textit{term-frequency} gives a measure of how often a given word appears across documents/corpus whereas, the \textit{inverse-term frequency} gives a measure of the importance of a word based on its rarity across documents. If a word appears in many documents frequently, it would have a high frequency and low rarity and thus, may not be really useful in the analysis. Term-frequency $tf$ in a document $d$ can be calculated by dividing the term's frequency by the total number of terms \cite{manning2008introduction}:

\[ tf = \frac{f_{t,d}}{\sum_{t'\in d} f_{t', d}} \]

Inverse-document frequency is calculated as $\log$ of total number of documents $D$ divided by the total number of documents containing the term $t$ \cite{manning2008introduction}:
\[
    idf = \log \left[ \frac{|D|}{\sum t\in d} \right]
\]

The combined $tf-idf$ score is given by the product of $tf$ and $idf$ values for the term $t$:
\[
    tf\mhyphen idf =  tf_{d} * idf_{t\in D}
\]

\subsection{Doc2Vec}
Document-to-vector (Doc2Vec), also known as \textit{paragraph-to-vector}, was introduced in 2014 as an extension to a similar word embedding model - Word2Vec \cite{Doc2vec}. In addition to learning word vectors from the corpus, Doc2Vec introduces an additional vector called \textit{paragraph vector} that represents unique features of each paragraph (such as the topic of the paragraph) in the corpus. Both the paragraph vectors and the context vector composed of word vectors are then used to predict the target word. This is similar to the Continuous-Bag-of-Words model (CBOW) in word2vec model \cite{word2vec}. The authors refer to this in their paper as \textit{Paragraph Vector-Distributed Memory}. In this model, the paragraph vectors act as a memory by predicting the target word using the context words taken from the paragraph and the paragraph id. Another variant of Doc2Vec is distributed bag of words (PV-DBOW), which is similar to \textit{skip-gram} of word2vec, wherein the target word is used to predict the surrounding context words.
\section{Methodology}
\label{sec:methodolgy}

This section discusses the employed methodology along with data collection for building predictive models to capture consequences of novel cyber attacks. 

\subsection{Data Collection}
As part of an ongoing research project, we collected 93 cyberattacks along with three descriptions explaining the attack compiled from publicly available security blogs and threat repositories such as: CWE \cite{CWE}, CAPEC \cite{CAPEC} and ATT\&CK \cite{ATTACK}. We then annotated these attacks with 50 non-technical consequence descriptions, which are written in simple language (i.e., layman terms) so they can be understood by security practitioners as well as CPS end-users. This is important for communication because various stakeholders who make decisions about cyberattack mitigation and strategy plans have little to no security knowledge. The 50 consequences were then grouped into 7 clusters based on the similarity between the consequences. The process involved an open card sort activity through which we recruited a number of participants to sort and group the 50 consequences into groups. Table \ref{tab:attacks} shows some of the attacks, consequences and their cluster number. The cluster numbers and their labels are described in Table \ref{tab:consequences}. We only list the cluster labels and not the 50 individual consequences for brevity.

\begin{figure}
    \centering
    \includegraphics[scale=0.55]{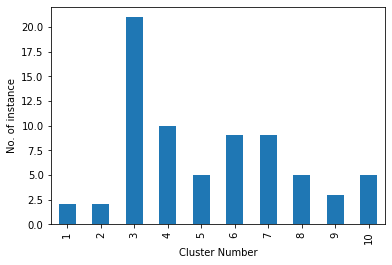}
    \caption{Number of Instances in cluster labels.}
    \label{fig:instances}
\end{figure}

\begin{table*}[]
\caption{Examples of the attacks, description, consequences and cluster number.}
\renewcommand{\arraystretch}{1.5}
\begin{tabular}{|p{2cm}|p{9cm}|p{4cm}|p{1cm}|}
\hline
\multicolumn{1}{|c|}{\textbf{Attack}} & \multicolumn{1}{c|}{\textbf{Description}} & \multicolumn{1}{c|}{\textbf{Consequence 1}} & \multicolumn{1}{c|}{\textbf{Cluster\#}} \\ \hline                    

Log Injection & An attacker can replicate log entries or inject malicious code by writing invalid user input. In most cases, this can be accomplished by entering certain characters. An attacker can also cause logs files to become unusable by corrupting the file format or inserting certain characters when rendered by systems that process these log files automatically. Additionally, attackers can also insert malicious code in the log files, which executes when the system parses the file. Corrupted log files can enable an attacker to hide their activities or can also make it look like some else performed the malicious activities \cite{loginjection}. & The cyber-attacker modified your computer files in order to hide their activities. & 4 \\ \hline

Webpage/URL Spoofing  &  Web spoofing is a type of attack that enables an attacker to change and observe the web pages sent to a victim’s browser. An attacker can also observe the information being entered by the victim into online-forms. This attack can happen even when there appears to be a secure connection to a website and can occur in any type of browser. The user is often oblivious to anything being out of place \cite{webspoofing}.  &  The cyber-attacker made you think an Internet site that the attacker created was a legitimate Internet site. & 5        \\ \hline                                                                              
DNS Spoofing & Domain Name Server (DNS) Spoofing, also known as DNS poisoning, is an attack wherein the incoming network traffic to a legitimate website is directed to a malicious website by compromising the vulnerabilities in a DNS server fulfilling the request. DNS spoofing poses several challenges such as data theft and as a result, major banking and e-commerce website are often the target for attackers. Additionally, redirecting to the fake website may result in downloading malware and Trojans onto the victim’s computers requesting the website. Simply cleaning the DNS cache alone is not a solution to this as the cache can get corrupted again. Flushing the DNS cache can be a potential solution to this problem \cite{dnsspoofing}.    & The cyber-attacker rerouted your Internet requests to a device that they control . & 6 \\ \hline

DLL Tampering   & Dynamic-linked libraries (DLL) are an important part of windows OS. It allows code reuse, modularization, and efficient memory utilization.  Thus, DLL are crucial to help programs load and faster. An attacker can tamper with the DLL files and disrupt the normal functioning of the programs and system \cite{dlltampering}.     & The cyber-attacker made your computer run software that your computer did not intend to run.  & 7       \\ \hline                                                                                 
TCP SYN Flood   & TCP protocol requires a three-way handshake to establish a connection. In TCP-SYN flood attacks, the attacker sends several new connection requests originating with different IP addresses. The server never receives an acknowledgment packet from the spoofed IP addresses and thousands of such connection requests cause the server to run of memory and it eventually crashes.  This prevents the legitimate users to connect to the server as well \cite{tcpudp}.   & The cyber-attacker caused your computer to crash.   & 7  \\ \hline

UDP Flood   & In UDP flood attack, a server is overloaded with UDP packets. UDP packets can be adjusted to a max size of 6500 bytes which can be used by attackers as a quick way to exhaust the server’s memory and bandwidth using just a few compromised systems to flood the server \cite{tcpudp}  &  The cyber-attacker caused your computer to run very slowly.  & 10    \\ \hline                                            
\end{tabular}
\label{tab:attacks}
\end{table*}

\begin{table}[!t]
\centering
\caption{List of Cluster Number and their Cluster Labels}
\begin{tabular}{|c|p{6.5cm}|}
\hline
\textbf{Cluster \#} & \multicolumn{1}{c|}{\textbf{Cluster Label}} \\ \hline
\textbf{Cluster 1}        & The attacker sent you emails that could lead to an attack if their request is granted                              \\ \hline

\textbf{Cluster 2}        & The attacker disrupted your access to your computer or the Internet.                                               \\ \hline

\textbf{Cluster 3}        & The attacker gained access to your computer or one of your online accounts.                                        \\ \hline

\textbf{Cluster 4}        & The attacker altered your computer or its contents to allow them to use it for their purposes without you knowing. \\ \hline

\textbf{Cluster 5}        & The attacker manipulated your use of or understanding about a website.                                             \\ \hline

\textbf{Cluster 6}        & The attacker changed or intercepted information that you have on the Internet.                                     \\ \hline

\textbf{Cluster 7}  & The attacker made your computer operate inefficiently or not at all.                                               \\ \hline

\textbf{Cluster 8}  & Cluster labels for 6, 5, and 3 \\ \hline

\textbf{Cluster 9}  & Cluster labels for 4, 7, and 2 \\ \hline

\textbf{Cluster 10}  & Cluster labels for 2, 7, and 7\\ \hline

\end{tabular}
\label{tab:consequences}
\end{table}

\subsection{Data Pre-processing}
Some of the attacks had consequences that belong to more than one cluster. To accommodate this, we created three additional cluster numbers and labels representing these combinations. For example, UDP flood attack in Table \ref{tab:attacks} had consequences belonging to clusters 2 and 7. We combined attacks having similar cluster numbers in any order (2,7,7 or 7,2,2) to be represented with cluster number 10. Thus, in addition to the 7 original cluster labels, we added 3 additional clusters labels representing a combination of consequences belonging to different clusters as shown in Table \ref{tab:consequences}. Thus, each attack ended up having a cluster number value between 1 to 10.

Fig. \ref{fig:instances} shows the overall distribution of instances per each cluster number. It is clear that some of the clusters had fewer than 2 data instances. In order to avoid a class imbalance  problem, we ended up ignoring clusters 1, 2 and 9 from the final dataset so as to avoid any bias during the training of the machine learning models.

\subsection{Text Cleaning}
The text data in the attack descriptions needs to be pre-processed before it can be useful for further processing. Text cleaning or pre-processing involves removing stopwords (such as articles), punctuation, digits, stemming and lemmatization. We used clean-text library\footnote{\url{https://pypi.org/project/clean-text/}} to convert the text into lower case and remove stopwords and punctuation. After this step, the text descriptions can be used for feature extraction.

\subsection{Feature Extraction}
We used \texttt{TfidfVectorizer} from sklearn library to generate the word-term matrix. We allowed for unigrams and bigrams to be included in the vocabulary and considered only words that have a document frequency of at least 2.
Additionally, we used gensim's \texttt{Doc2Vec} library, to get feature vectors for attack descriptions. We used a vocabulary size of 300 based on preliminary experiments.

\subsection{Experimental Setup}

The final dataset consists of 72 cyberattacks after excluding physical attacks, such as using a USB Killer, as we wanted to focus only on attacks in the cyber-space. The dataset includes the attack's name, attack description, technical and non-technical consequences and the cluster number and label. The entire dataset was divided into training and test sets with a 70\% and 30\% split, respectively. Additionally, we used stratified samples based on the number of samples per class to account for the class imbalance. 

We used both tf-idf and Doc2Vec features described in Section \ref{sec:background} to compare results across different machine learning models. For the tf-idf method, we included only words that have a minimum document frequency of 2 and max document frequency of 0.98. We also allowed for unigrams and bigrams to be included in the tf-idf matrix.
For Doc2Vec, we used the distributed bag of words model with vocabulary size of 300 and learning rate 0.065. The Doc2Vec model was trained for 50 epochs. We used four machine learning models from the sklearn python library: \textit{LinearSVC, Logistic Regression, Multinomial Naive Bayes, Random forest}, and \textit{Multilayer Perceptron}. The results are described in the next section.

\section{Results}
\label{sec:results}
Table \ref{tab:results} shows performance of the various models for both tf-idf and Doc2Vec methods. After text cleaning, each text description was converted to the appropriate format that the underlying library required. For example, in the case of tf-idf, the text descriptions were converted to a matrix with rows representing the documents and columns containing the words. Similarly, for Doc2Vec, the training and testing documents were tokenized and tagged with a label using the \texttt{TaggedDocument} class from Doc2Vec library. The features were then used to train the ML models to learn and predict the cluster number. This problem is an instance of the general  multi-class classification wherein, the target has multiple class labels instead of binary labels.

\begin{table}[!h]
\caption{Results of the classification models for predicting Cluster number.}
\centering
\begin{tabular}{|l|r|r|r|r|}
\hline
\multicolumn{5}{|c|}{\textbf{tf-idf}}\\ \hline
\textbf{Model}                & \textbf{Accuracy}  & \textbf{Precision} & \textbf{Recall}  & \textbf{F1-score}  \\ \hline
Logistic Regression    & 0.60 &  0.45 & 0.60 & 0.50       \\ \hline
Linear SVC    & \textbf{0.60} &  0.49 &  0.60 & \textbf{0.53}               \\  \hline
Multilayer Perceptron & 0.60 & 0.51 & 0.60 & 0.50          \\  \hline
MultinomialNB          &  0.40  & 0.28 &  0.40 & 0.26       \\ \hline
RandomForestClassifier & 0.40 & 0.16 & 0.40 &  0.23         \\ \hline
\multicolumn{5}{|c|}{\textbf{Doc2Vec}}\\ \hline
Logistic Regression    & 0.48 &  0.53 & 0.48 & 0.46       \\ \hline
Linear SVC    & \textbf{0.57} &  0.57 &  0.57 & \textbf{0.53}              \\  \hline
Multilayer Perceptron & 0.38 & 0.43 & 0.38 & 0.37          \\  \hline
GaussianNB          &  0.32  & 0.23 &  0.32 & 0.25       \\ \hline
RandomForestClassifier & 0.42 & 0.37 & 0.42 &  0.35         \\ \hline
\end{tabular}
\label{tab:results}
\end{table}

{\it 1) tf-idf.} In the case of the tf-idf method, the \textit{Linear-SVC} method is selected as the best performing model over \textit{Logistic Regression} and \textit{Multilayer Perceptron} as it has a higher F1-score compared to the other two models. Surprisingly, the worst performing model is the random forest classifier. This is against the supporting literature in prediction and classification modeling in some other security related topics such as fake reviews identification \cite{DBLP:conf/compsac/Gutierrez-Espinoza20} and zero-day malware detection \cite{DBLP:conf/bigdataconf/AbriSKSN19}. A possible reason might be due to the diversity of cyber security descriptions and thus difficulty in building decision trees when conducting ensemble prediction models.  

{\it 2) Doc2Vec.} Similarly, for the Doc2Vec method, \textit{LinearSVC} outperformed the other classifiers.  \textit{LinearSVC} achieved an accuracy score of 57\% and an F1-score of 53\%. While the performance of random forest classifier has not been improved, the performance of the other classifiers such as multilayer perceptron and Gaussian NB has been degraded. Here, we studied simple multilayer perceptron. It would be intriguing to investigate the performance of deeper versions of Convolutional \cite{DBLP:conf/compsac/Tavakoli20} or Recurrent Neural Networks \cite{DBLP:conf/bigdataconf/Siami-NaminiTN19} and optimize the prediction models accordingly. 

As Table \ref{tab:results} indicates, \textit{Linear-SVC} model had the best performance for both the tf-idf and Doc2Vec methods with 0.6 and 0.57 accuracy, respectively. It is worth noting that tf-idf and Doc2Vec are different in that tf-odf considers the frequency of the terms in the document (i.e., email); whereas, Doc2Vec focuses on the semantics of documents. Given that we obtain a slightly better result using tf-idf than Doc2Vec, it may indicate that for phishing detection, email embedding through extracting frequency features might provide a better classification. The results of predicting consequences of attacks are promising and it indicates that with some additional hypertuning and optimization it is possible to hit a higher accuracy.

\section{Conclusion and Future Work}
\label{sec:conclusion}

In this paper, we employed machine learning models to predict the  consequences of cyber attacks using two popular word embedding methods, that is, tf-idf and Doc2Vec. LinearSVC achieved the best performance for both the cases, which is consistent with past research that demonstrated LinearSVC to be well suited for multiclass-classification problems in natural language processing tasks. Although, the best accuracy obtained was only 60\%, this could be attributed to the small data sample with an unequal number of samples per cluster label. We tried to accommodate for this problem by using stratified sampling during train-test splitting. A stratified approach balances the samples in the train-test split such that data instances for no single class can bias the ML models.  

In future work, we would like to explore complex features using other word-embeddings, such as Word2Vec, to evaluate if the prediction scores can be improved. It would be also to reduce the dimensionality of the output of Doc2Vec using feature reductions techniques such as principal component analysis and encoder-decoder \cite{DBLP:journals/corr/abs-2004-07296}. These emerging techniques have shown great performance in recent research work. Additionally, we would also like to explore whether we can use other natural language processing techniques to directly predict the actual consequence itself instead of the cluster number.

    \vspace{-0.2in}
\section*{Acknowledgment}
This research work is supported by National Science Foundation (NSF) under Grant No: 1564293.

\bibliographystyle{IEEEtran}
\bibliography{main}
\end{document}